\documentclass[aps,preprint,showpacs]{revtex4}
\usepackage{epsfig}
\usepackage{amsmath}
\usepackage{amssymb}
\usepackage{amsfonts}
\usepackage{enumerate}

\begin{document}
\title{The inequality of charge and 
spin diffusion coefficients}
\author{Sandipan Pramanik and Supriyo Bandyopadhyay \footnote{Corresponding 
author. E-mail:
sbandy@vcu.edu}}
\affiliation{Department of Electrical and Computer Engineering, Virginia
Commonwealth University, Richmond, VA 23284, USA}
\author{Marc Cahay}
\affiliation{Department of Electrical and Computer Engineering and Computer
Science,
University of Cincinnati, Cincinnati, OH 45221, USA}

\bigskip

\begin{abstract}

Since spin and charge are both carried by electrons (or holes) in a solid, it is 
natural to assume that charge and 
spin diffusion coefficients will be the same. Drift-diffusion models of spin 
transport typically assume so. Here, we show analytically that the two diffusion 
coefficients can be vastly 
different in  quantum wires. Although we do not consider quantum wells or bulk 
systems, 
it is likely that the two coefficients will be different in those systems as 
well. 
Thus, it is important to distinguish between them 
in transport models, particularly those applied to quantum wire based devices.

\end{abstract}

\maketitle
\pagebreak

In the drift-diffusion model of spin transport, it is customary to assume 
 that the {\it same} diffusion 
coefficient `$D$' describes charge and spin diffusion. This assumption 
is commonplace in the literature (see, for example, refs. 
\cite{zhang,mishchenko,burkov,saikin,pershin}). Ref. \cite{malshukov} considers 
a two dimensional system with different spin and charge diffusion coefficients 
but 
ultimately 
assumes that the bare spin diffusion coefficient is the same as the charge
diffusion coefficient. Ref. \cite{flatte} also examines this issue, and based on 
an 
heuristic assumption that spin transport is analogous to bipolar charge 
transport, reaches the conclusion that the two diffusion coefficients are equal 
as long as the populations of upspin and downspin carriers are {\it equal}. In 
spin polarized transport, the two populations are unequal by definition. 
Therefore, it is imperative to examine if these two diffusion coefficients are 
still equal in spin polarized transport, and if not, then how unequal they can 
be. In this paper, we show that these two diffusion coefficients can be vastly 
different in quantum wires. Although we do not consider quantum wells and bulk 
systems, there is no reason to believe {\it apriori} that even in those systems, 
the two diffusion 
coefficients 
will be equal. 

We first consider a narrow semiconductor quantum 
wire 
where only the lowest subband is occupied by carriers at all times. 
All higher subbands are unoccupied. We will assume that there are 
Rashba \cite{rashba} and Dresselhaus \cite{dresselhaus} spin orbit interactions 
in the wire, but no external magnetic field to cause spin mixing 
\cite{cahay}. 
In that case, we can ignore the Elliott-Yafet spin relaxation mechanism 
\cite{elliott} since 
it will be very weak. Spin relaxation 
via hyperfine interaction with nuclear spins, or via the Bir-Aronov-Pikus 
mechanism \cite{bap}, is also typically
very weak in semiconductor quantum wires with only one kind of carriers 
(electrons or holes, but not both). Therefore, the only spin relaxation 
mechanism that is important is the D'yakonov-Perel'' relaxation \cite{dp}.

In the single channeled quantum wire, we will prove two remarkable results: (i) spin will 
relax in 
{\it time} (i.e. the spin relaxation time $\tau_s$ will be finite), but it will not relax in {\it space}
(i.e. the spin relaxation length $L_s$ will be infinite), and (ii) if the 
drift-diffusion model is valid in this system (this model relates $L_s$ and 
$\tau_s$ as $L_s$ = $\sqrt{D_s \tau_s}$), then we must conclude that 
the spin diffusion coefficient $D_s$ is  
infinite. However, since there is scattering in the system, the charge diffusion 
coefficient $D_c$ must be finite. Therefore, the two diffusion coefficients are completely different.
This is an extreme case, but even in less extreme cases (multi-channeled quantum wires), these two 
coefficients can be very 
different.
Below, we provide an analytical proof for the single channeled quantum wire case.

Consider an ensemble of electrons injected in a quantum wire at time $t=0$ from 
the end $x=0$ as shown in Fig. 1. Only the lowest subband is occupied in the 
wire at all times. There is an electric field $E_x$ driving charge transport, 
and there is also a transverse electric field $E_y$  breaking structural 
inversion symmetry, thereby causing a Rashba spin orbit interaction 
\cite{rashba}. We will assume that the quantum wire axis is along the [100] 
crystallographic direction and that there is crystallographic inversion 
asymmetry along this direction giving rise to Dresselhaus spin-orbit interaction 
\cite{dresselhaus}. We choose this system because it is the simplest and 
includes 
the two major types of spin orbit interactions found in semiconductor nanostructures, namely the 
Rashba
and the Dresselhaus interactions. Ref. 
\cite{pershin} has considered this system within the framework of the  
drift-diffusion model and shown that there is a single time constant describing 
spin relaxation. In contrast, spin relaxation in a two-dimensional system 
(quantum well) may be described by more than one time constant \cite{pershin}.

For illustration purposes, we will assume hypothetically that the spin injection 
efficiency is 100\%, so that at $x=0$, all electrons are spin polarized along 
some particular, though arbitrary, direction $\hat\eta_0$ in space. Their 
injection velocities are not necessarily the same (in fact, they will be drawn from the 
Fermi-Dirac distribution in the contact). We are interested in finding 
out how the net spin polarization of the ensemble ($|\langle\bf S\rangle |$) 
decays in time or space due to the D'yakonov-Perel' process. 

In the quantum wire, the electrons experience various momentum relaxing 
scattering events. Between successive scattering events, they undergo free 
flight and during this time, their spins precess about a velocity-dependent 
pseudo-magnetic field ${\bf B_{so}}(v_x)$ caused by Rashba and Dresselhaus 
spin-orbit interactions. This magnetic field can be shown to be spin-independent.

The spin precession of every {\it single} electron occurs according to the well-known Larmor equation:
\begin{equation}
{{d {\bf S}}\over{dt}} = {\bf \Omega }(v_x)  \times {\bf S},
\label{precession}
\end{equation}
where ${\bf S}$ is the spin polarization vector of the electron and ${\bf \Omega} (v_x)$ is a 
vector whose magnitude is the angular frequency of spin precession. It is  
related 
to ${\bf B_{so}}(v_x)$ as ${\bf \Omega }(v_x)  = 
(g \mu_B/\hbar) {\bf B_{so}}(v_x)$, where $g$ is the Land\'e g-factor in the material and $\mu_B$ is the Bohr magneton. This is actually the well-known 
equation for Larmor spin precession and can be derived rigorously from the Ehrenfest Theorem of quantum mechanics.

The vector  ${\bf \Omega }(v_x) $ has two contributions due to Dresselhaus and 
Rashba interactions:
\begin{equation}
{\bf \Omega}(v_x) = {\bf \Omega_D} (v_x) + {\bf \Omega_R} (v_x)
,
\end{equation}
where the first term is the Dresselhaus and the second term is the Rashba 
contribution. These two contributions are given by
\begin{eqnarray}
{\bf \Omega_D} (v_x) & = & {{2 m^* a_{42}}\over{ \hbar^2}}\left 
[ 
\left({{ 
\pi}\over{W_y}} \right )^2 - \left({{ \pi}\over{W_z}} \right )^2 \right ] v_x 
\hat{x} = \zeta_{D0} v_x \hat{x} \nonumber \\
{\bf \Omega_R }(v_x) & = & {{2 m^* a_{46}}\over{ \hbar^2}} E_y v_x 
\hat{z} 
= \zeta_{R0} v_x \hat{z}
,
\label{contributions}
\end{eqnarray}
where $W_z, W_y$ are the transverse 
dimensions of the wire,  $a_{42}$ and $a_{46}$ are material constants, $\hat{x}$ 
is the unit vector along the x-direction and $\hat{z}$ is the unit vector along 
the z-direction.

Note that the vector ${\bf \Omega}$ lies in the x-z plane and subtends an angle 
$\theta$ with the $\pm$x-axis (quantum wire axis) given by 
\begin{equation}
\theta = arctan\left [ {{\zeta_{R0}}\over{\zeta_{D0}}} \right ] = arctan \left [ 
{{a_{46} E_y}\over{a_{42} \left \{
\left({{ 
\pi}\over{W_y}} \right )^2 - \left({{ \pi}\over{W_z}} \right )^2 \right \} }} 
\right ]
.
\end{equation}

Note also that since $\theta$ is independent of $v_x$, the {\it axis} (but not 
the magnitude) of both ${\bf \Omega}$ and  ${\bf B_{so}}$ is independent of electron 
velocity. Therefore, every electron, regardless of its velocity, precesses about 
the {\it same} axis, as long as only one subband is occupied. The direction of 
precession (clockwise or counter-clockwise) depends on the sign of the velocity 
and therefore can change if the velocity changes sign, but the precession {\it axis} 
remains unchanged. However, the precession frequency depends on the velocity and 
is therefore different for different electrons as long as  there is a spread in 
their velocities caused by varying injection conditions or random scattering. As 
a result, at any given instant of time $t$ = $t_0$, the spins of different 
electrons will be pointing in different directions because they have precessed 
by different angles since the initial injection.
Consequently, when we ensemble average over all electrons, the quantity $|\langle\bf 
S\rangle |$ decays in time, leading to spin relaxation {\it in time}.

To show this more clearly, we start from Equation (\ref{precession}) describing the 
spin precession of any one arbitrary electron: 
\begin{eqnarray}
{{d {\bf S}}\over{dt}} & = & \hat{x}{{d S_x}\over{dt}} + \hat{y}{{d 
S_y}\over{dt}} + \hat{z}{{d S_z}\over{dt}}
\nonumber \\
& = & {\bf \Omega }(v_x)  \times {\bf S} \nonumber \\
& = & det \left [ \begin{array}{ccc}
\hat{x} & \hat{y} & \hat{z} \\
 \Omega_D (v_x) & 0 & \Omega_R ( v_x) \\
 S_x & S_y & S_z \\
\end{array}
\right ] \nonumber \\
& = & -\hat{x} \left [ \Omega_R (v_x) S_y \right ] - \hat{y} \left 
[\Omega_D 
( v_x) S_z - \Omega_R (v_x) S_x \right ] + \hat{z} \left [ \Omega_D 
(v_x) S_y \right ]
,
\end{eqnarray}
where $S_n$ is the spin component along the $n$-axis of that arbitrary electron.

Equating each Cartesian component separately, 
we get:
\begin{eqnarray}
{{d S_x}\over{dt}} & = &  -\zeta_{R0}v_xS_y, 
\nonumber \\
{{d S_y}\over{dt}} & = &  = \zeta_{R0}v_xS_x - 
\zeta_{D0}v_x S_z, \nonumber \\
{{d S_z}\over{dt}} & = & \zeta_{D0}v_xS_y.
\label{components} 
\end{eqnarray}

If every electron in an ensemble had the {\it same} $v_x$ at {\it every instant of time} 
(no dispersion in velocity), then the last equation 
tells us that every electron would have the exact same spin components $S_x$, $S_y$ and $S_z$
at any instant of time as long as they were all injected at time $t$ = 0 
with the same spin polarization. In that case, we could replace 
$S_n$ in the last equation by the ensemble averaged value $<S_n>$ over the entire ensemble, so that
\begin{eqnarray}
{{d |<S>|^2}\over{dt}} & = & {{d <S_x>^2}\over{dt}} + {{d <S_y>^2}\over{dt}} + 
{{d <S_z>^2}\over{dt}} \nonumber \\
& = & 2 <S_x>{{d <S_x>}\over{dt}} + 2 <S_y>{{d <S_y>}\over{dt}} + 2 <S_z>{{d 
<S_z>}\over{dt}} \nonumber \\
& = &  -2\zeta_{R0}v_x<S_y> <S_x> + 2\zeta_{R0}v_x<S_x><S_y> - 
2\zeta_{D0}v_x <S_z><S_y> \nonumber \\
&& + 2\zeta_{D0}v_x<S_y><S_z> \nonumber \\
& = & 0.
\label{eq-7}
\end{eqnarray}

In that case, $|<S>|$ will not decay in time and there will be no 
D'yakonov-Perel' spin relaxation in time. However, if $v_x$ is {\it different} 
for different electrons either due to different injection conditions, or because 
of scattering, then we cannot replace $S_n$ with $<S_n>$ in Equation 
(\ref{components}). As a result, Equation (\ref{eq-7}) will not hold, so that $d|<S>|/dt$ $\neq$ 0, and there will be a 
D'yakonov-Perel' relaxation in time. As a result, the spin relaxation time 
$\tau_s$ will be {\it finite}.

Next, let us consider D'yakonov-Perel' spin relaxation in {\it space}. From 
Equation 
(\ref{components}), we obtain (using the chain rule of differentiation)
\begin{eqnarray}
{{d S_x}\over{dx}}{{d x}\over{dt}} & = & {{d S_x}\over{dx}}v_x = 
-\zeta_{R0}v_xS_y ,
\nonumber \\
{{d S_y}\over{dx}}{{d x}\over{dt}} & = & {{d S_y}\over{dx}}v_x = 
\zeta_{R0}v_xS_x - 
\zeta_{D0}v_x S_z , \nonumber \\
{{d S_z}\over{dx}}{{d x}\over{dt}} & = & {{d S_z}\over{dx}}v_x = 
\zeta_{D0}v_xS_y
.
\label{components1} 
\end{eqnarray}

The above equation shows that the spatial rates $dS_n/dx$ are {\it independent} of velocity. 
This is a remarkable result with remarkable consequence. It tells us that 
even if different electrons 
have different velocities, as long as they were 
all injected with the same spin polarization at $x$ = 0, they will all have the exact same 
spin polarization at any arbitrary location $x$ = $X_0$!
That is, every electron's spin at $x$ = $X_0$ is pointing in exactly the 
same direction.
Therefore, we can always replace $S_n$ in the above equation by its ensemble 
averaged value $<S_n>$ whether or not there is scattering causing a spread in 
the electron velocity between different members of the ensemble. 
 Consequently,
\begin{eqnarray}
{{d |<S>|^2}\over{dx}} & = & {{d <S_x>^2}\over{dx}} + {{d <S_y>^2}\over{dx}} + 
{{d <S_z>^2}\over{dx}} \nonumber \\
& = & 2 <S_x>{{d <S_x>}\over{dx}} + 2 <S_y>{{d <S_y>}\over{dx}} + 2 <S_z>{{d 
<S_z>}\over{dx}} \nonumber \\
& = &  -2\zeta_{R0}<S_y> <S_x> + 2\zeta_{R0}<S_x><S_y> - 
2\zeta_{D0} <S_z><S_y> \nonumber \\
&& + 2\zeta_{D0}<S_y><S_z> \nonumber \\
& = & 0
.
\end{eqnarray}
Thus, there is never any D'yakonov-Perel relaxation in {\it space} as long as a 
single subband is occupied. Therefore, the spin relaxation length
$L_s$ is {\it infinite}. This is true whether or not there is scattering. 

The above result has been confirmed independently with a many-particle Monte Carlo simulation
of spin transport in a single channeled quantum wire \cite{pramanik-ieee}.
Here, we have provided an analytical proof.

The foregoing analysis also shows that in a quantum wire with single subband occupancy and 
D'yakonov-Perel' as the only spin relaxation mechanism, there is a fundamental 
difference between spin relaxation in {\it time} and spin relaxation in {\it 
space}. Spin can relax in time while not relaxing in space. The physical origin 
of this difference is explained below:

From Equation (\ref{contributions}), we see that the precession frequency 
for any arbitrary electron is given by
\begin{equation}
{{d \phi(t)}\over{dt}}  = | {\bf \Omega}|(t) = \sqrt{\zeta_{D0}^2 + \zeta_{R0}^2}v_x(t) = \zeta_0 
v_x(t)
,
\end{equation}
where $\phi(t)$ is the angle by which the electron's spin precesses in time $t$.

If all electrons are injected with the same spin polarization at time $t$ = 0, 
then the angle by which any given electron's spin has precessed at time $t$ = 
$t_0$ is
\begin{equation}
\phi(t_0) =  \zeta_0 \int_0^{t_0} v_x(t) dt = \zeta_0 [x(t_0) - x(0)] = \zeta 
d_0
,
\end{equation}
where $d_0$ is the distance between the location of the electron at time $t_0$ 
and the point of injection. Obviously $d_0$ is {\it history-dependent}, because 
different electrons with different injection velocities and/or scattering 
histories would traverse different distances in  time $t_0$. Consequently, if we 
denote the angle by which the $n$-th electron's spin has precessed in time $t_0$ 
 as $\phi_n(t_0)$, then $\phi_1(t_0) \neq \phi_2(t_0) \neq ... \phi_m(t_0)$. As 
a result, if we 
take a snapshot at  $t_0$, we will find that the spin polarization vectors of 
different electrons are pointing in different directions. Therefore, ensemble 
averaged spin at $t_0$ is less than what it was at time $t$ = 0. Consequently, 
spin depolarizes with time leading to {\it temporal} D'yakonov-Perel' relaxation.

The spatial rate of precession, on the other hand, is obtained as
\begin{eqnarray}
{{d \phi(t)}\over{dt}} & = & {{d \phi(x)}\over{dx}}{{d x}\over{dt}} = {{d 
\phi(x)}\over{dx}} v_x(t) = \zeta_0 v_x(t) \nonumber \\
{{d \phi(x)}\over{dx}} & = & \zeta_0
.
\end{eqnarray}
Therefore, the angle by which any given electron's spin has precessed when it 
arrives at a location $x$ = $X_0$ is 
\begin{equation}
\phi(X_0) = \int_0^{X_0} {{d \phi(x)}\over{dx}} dx = \zeta_0 \int_0^{X_0} dx = 
\zeta_0 X_0
.
\end{equation}
This angle is obviously {\it history-independent} since it depends only on the 
coordinate $X_0$ which is the same for all electrons at location $X_0$, 
regardless of how and when they arrived at that location. In fact, an electron 
may have 
visited the location $X_0$ earlier, gone past it, and then scattered back to 
$X_0$. Or it may have arrived at $X_0$ for the first time. It does not matter. 
The angle by which an electron's spin has 
precessed when it is located at $X_0$ is a constant independent of past history. 
Therefore, if all electrons were injected with
their spins exactly parallel to each other at $x$ = 0, then every single 
electron at $x$ = $X_0$ has its spin polarization vector pointing in the {\it 
same} direction as every other electron, and the ensemble averaged {\it magnitude} of spin at $x$ = $X_0$ is the same as 
that at $x$ = 0. Consequently, spin does not depolarize in space and there is no 
D'yakonov-Perel spin relaxation in space, unlike time.

Since spin relaxes in time but not in space, the relaxation time ($\tau_s$) is 
finite whereas the relaxation length ($L_s$) is infinite. According to the drift-diffusion model, 
these two quantities 
are always related in steady state as \cite{saikin}
\begin{equation}
L_s = \sqrt{D_s \tau_s}
,
\label{relation_spin}
\end{equation}
where $D_s$ is the spin diffusion coefficient. Note that the quantities $L_s$ 
$D_s$ and $\tau_s$ are spin transport constants. As such, they are independent 
of both space and time.

Since $L_s$ is infinite while $\tau_s$ is finite, the only way the above
equation can be satisfied is if the steady-state spin diffusion coefficient 
$D_s$ is 
infinite. But the steady state diffusion coefficient $D_c$ associated with 
charge transport 
is 
certainly finite since we have frequent momentum relaxing scattering in our 
system. Therefore, there must be two very {\it different} diffusion coefficients 
$D_s$ and $D_c$ 
associated with spin and charge diffusion. This completes our analytical proof 
that $D_c \neq D_s$.

Two final questions remain regarding the generality of the above result. First, 
is it only 
valid for the extreme case of a quantum wire with single subband occupancy (single channeled 
transport) and second, is it only true for Dyakonov-Perel relaxation? 
We cannot treat the case of multi-channeled transport analytically, but we have 
examined 
that case numerically using Monte Carlo simulation in both space 
\cite{pramanik_apl} and time 
\cite{pramanik_prb}. We studied spin transport in a GaAs quantum wire of cross 
section 30 nm 
$\times$ 4 nm, where multiple subbands are occupied and Dyakonov-Perel' 
relaxation does occur in both 
time and space. At a lattice temperature of 77 K and a driving electric field 
$E_x$ = 2 kV/cm,
the value of $L_s$  extracted from that study  is $\sim 10 ~ \mu m$ while the 
value of $\tau_s\sim 1 ~ {\text{nsec}}$. This yields $D_s\sim 10^3 
~{\text{cm}}^2/{\text{s}}$ (from Equation (\ref{relation_spin})) which 
is still several orders of magnitude higher than the charge diffusion 
coefficient 
$D_c$ in the same quantum wire calculated under the same conditions 
\cite{telang1,telang2}.
Thus 
$D_s \neq D_c$, even in multi-channeled transport, and the two quantities can be 
vastly 
different. 

Finally, what if we include other modes of spin relaxation, such as 
Elliott-Yafet \cite{elliott}?  If Elliott-Yafet is the dominant mode, then spin relaxation is 
intimately connected with momentum relaxation. In that case, the charge 
diffusion constant, determined by momentum relaxing scattering, and spin 
diffusion constant may not be as unequal. Nontheless, there is no reason to 
assume apriori that the two diffusion coefficients are exactly equal even in 
this case. A rigorous Monte Carlo simulation (based on random walk model) recently carried out by us has shown that the two diffusion coefficients, in general, are vastly different. How different they are depends on the details of the scattering processes that relax momentum and spin \cite{wan}.

In conclusion, we have shown that in quantum wires,
the spin and charge diffusion coefficients are vastly different. Although we 
have
not examined quantum wells and bulk systems in this study, there is no reason
to pre-suppose that the charge and spin diffusion coefficients will be equal 
in these systems either. Thus, it is important to distinguish between these 
two diffusion coefficients in solid state systems.



\clearpage

{\bf{Figure captions :}}\\

\begin{figure}
\epsfig{file=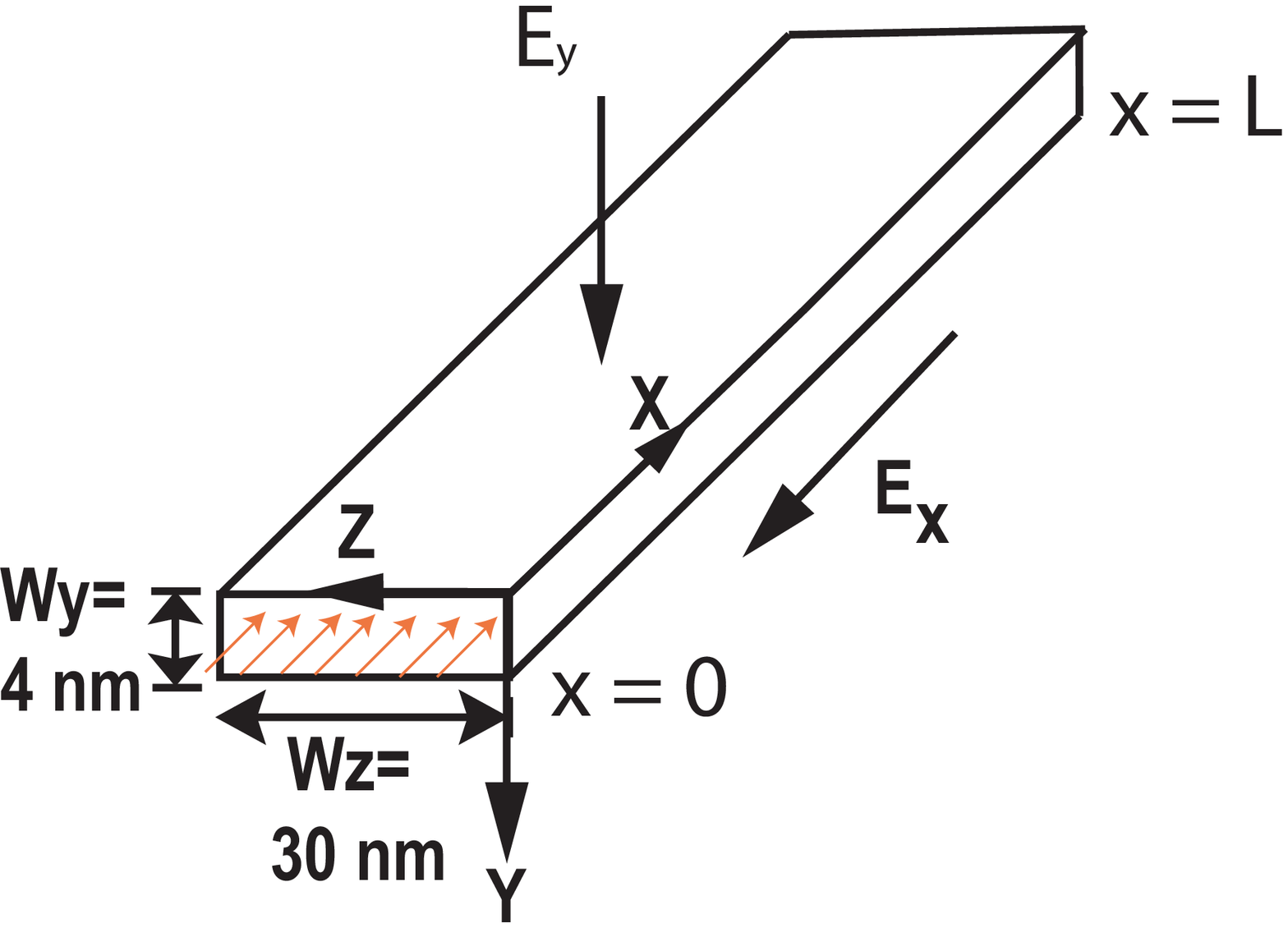, width=6in}
\caption{Pramanik et al.}
\label{model}
\end{figure}

Figure {\ref{model}}. A quantum wire structure of with rectangular 
cross section. A top gate (not drawn) applies a symmetry breaking electric field 
$E_y$ to induce Rashba interaction. A battery (not drawn) applies an electric 
field $-E_x\hat x$, $E_x>0$, along the channel. Spin polarized electrons are 
injected at $x=0$. These electrons 
travel along $\hat x$ and may gradually lose their initial spin polarization. We 
investigate the spin 
depolarization of these electrons in time domain as well as in space domain.

\clearpage

\end{document}